\documentclass{ws-ijmpcs}

\newcommand{\be}{\begin{equation}}
\newcommand{\ee}{\end{equation}}
\newcommand{\bea}{\begin{eqnarray}}
\newcommand{\eea}{\end{eqnarray}}

\newcommand{\bfk}{\mbox{\boldmath $k$}}
\newcommand{\bfq}{\mbox{\boldmath $q$}}

\newcommand{\pup}{p^\uparrow}

\newcommand{\kt}{k_\perp}
\newcommand{\bkt}{\mbox{\boldmath$k_\perp$}}
\def\lsim{\mathrel{\rlap{\lower4pt\hbox{\hskip1pt$\sim$}}\raise1pt\hbox{$<$}}}
\def\gsim{\mathrel{\rlap{\lower4pt\hbox{\hskip1pt$\sim$}}\raise1pt\hbox{$>$}}}

\begin{document}

\markboth{Asmita Mukherjee}
{Sivers Asymmetry in $e+p^\uparrow \rightarrow e+J/\psi+X$}

\catchline{}{}{}{}{}
%
%%%%%%%%%%%%%%%%%%%%%%%%%%%%%%%%%%%%%%%%%%%%%%%%%%%%%%%%%%%%%%%%%%%%

\title{SIVERS ASYMMETRY IN  $e+p^\uparrow \rightarrow e+J/\psi+X$ 
}

\author{ASMITA MUKHERJEE}

\address{Department of Physics, Indian Institute of Technology Bombay,\\
Powai, Mumbai 400076, India}

\maketitle

%\begin{history}
%\received{Day Month Year}
%\revised{Day Month Year}
%\end{history}

\begin{abstract}
A recent investigation of the single spin asymmetry (SSA) in
low virtuality electroproduction/photoproduction  of $J/\psi$ in color 
evaporation model is presented. It is shown that the asymmetry is sizable and 
can be used as a probe for the still  unknown gluon Sivers function.

\keywords{single spin asymmetry, Sivers function, TMD}
\end{abstract}

\ccode{PACS numbers: 13.88.+e, 13.60.-r, 14.40.Lb}

\section{Introduction}	

Large single spin asymmetries (SSA) observed when an unpolarized
beam of electrons or protons is scattered off a transversely polarized
target can be explained  with  the inclusion of ${\bkt}$ dependence in parton
distribution functions (pdf's) and fragmentation functions(ff's)
\cite{tmd-fact1}. In recent years, there has been a lot of interest in 
investigations of transverse single spin asymmetries
in high energy QCD processes as they provide information about spin and
orbital angular momentum carried by the quarks and gluons.
Large SSA's have been measured in  pion production at Fermilab
\cite{AdamsBravar1991} as well as 
at BNL-RHIC in $p p^\uparrow $ collisions~\cite{KruegerAllogower1999}.
SSA's have also been observed by HERMES \cite{Hermes} and COMPASS
\cite{Compass} collaborations, 
in polarized semi-inclusive deep inelastic scattering. 
The magnitude of the observed asymmetries have been found to be larger than
what is expected from perturbative QCD \cite{alesio-review}.
It is possible to
explain this large asymmetry in terms of transverse momentum depenent (TMD)
parton distribution functions.
 One is led  to a generalized factorization formula called
TMD factorization \cite{Sivers1990,tmd-fact2}, which  in some
processes has been proved at leading twist and leading order \cite{fact} and
has been argued to hold at all orders.  
One of the TMD functions is the  Sivers function which
describes the probability of finding an
unpolarized parton inside a transversely polarized hadron.
Here, we propose charmonium production as a probe to investigate the
Sivers function and estimate SSA in photoproduction (low   
virtuality electroproduction) of 
charmonium in scattering of  electrons off transversely polarized protons.
At leading order (LO), this  receives contribution only from a 
single partonic subprocess $\gamma g \rightarrow c {\bar c}$ . Hence,  SSA
in $e + p^\uparrow \rightarrow e +J/\psi +X$, if observed, can be used as a
clean probe of gluon Sivers function. In addition, charmonium production
mechanism can also have implications for this SSA and therefore, its 
study can help to understand the production mechanism for charmonium.

%Please provide a shortened runninghead (not more than eight words) for
%the title of your paper. This will appear on the top right-hand side
%of your paper.

%%%%%%%%%%%%%%%%%%%%%%%%%%%%%%%%%%%%%%%%%%%%%%%%%%%%%%%%%%%%%%%%%%%%%%%
\section{Sivers Asymmetry in $J/\psi$ Production : Formulas}
%%%%%%%%%%%%%%%%%%%%%%%%%%%%%%%%%%%%%%%%%%%%%%%%%%%%%%%%%%%%%%%%%%%%%%%

There are several models for charmonium production. 
In the color singlet model \cite{sing} 
the cross section for charmonium production is factorized into a short
distance part for $c {\bar c}$ pair production 
calculable in perturbation theory and a non-perturbative matrix element
for the formation of a bound state, which is produced in a color singlet
state. A more recent model of charmonium production is the color octet
model \cite{octet}. This is based on a factorization approach in
non-relativistic QCD (NRQCD) and it allows $c {\bar c}$ pairs
to be produced in color octet states. Here again, one 
requires knowledge of the nonperturbative colour octet matrix elements,
which are determined  through fits to the data on charmonium production.
On the other hand, in color evaporation model (CEM), which we use 
in this work, 
\cite{hal},\cite{fri} a statistical treatment of color is made
and the probability of finding a specific quarkonium state is assumed to be 
independent of the color of heavy quark pair. 
We have used Weizsacker-Williams (WW) equivalent photon approximation for the
photon distribution of the electron \cite{wwf1,wwf2}, to
calculate the cross section for the process 
$e+\pup\rightarrow e+ J/\psi+X$ at low virtuality of the photon. The
underlying  partonic process at
LO is $\gamma g\rightarrow c\bar{c}$ and therefore,
the only $\kt$ dependent pdf appearing is the gluon Sivers function.
For a complete calculation of photoproduction of $J/\psi$ one has to
consider higher order contributions and also the resolved photon
contributions \cite{ce2}. Also the gauge links or 
Wilson lines present in the TMD
distributions are important at higher order \cite{feng}.

According to CEM, the cross section for charmonium production is
proportional to the
rate of production of $c\bar{c}$ pair integrated over the mass range $2m_c$
to $2m_D$ 
\be
\sigma=\frac{1}{9}\int_{2m_c}^{2m_D} dM \frac{d\sigma_{c\bar{c}}}
{dM}
\ee 
where $m_c$ is the charm quark mass and $2m_D$ is the $D\bar{D}$ threshold,
$M^2$ is the squared invariant mass of the $c {\bar c}$ pair.
The WW distribution function of the photon in the electron given by
\cite{ww},
\bea
f_{\gamma/e}(r,E)=\frac{\alpha}{\pi} \{\frac{1+(1-r)^2}{r}
\left(ln\frac{E}{m}-\frac{1}{2}\right)
+\frac{r}{2}\left[ln\left(\frac{2}{r}-2\right)+1\right] \nonumber \\
+\frac{(2-r)^2}{2r}ln\left(\frac{2-2r}{2-r}\right) \}.
\label{ww-function}
\eea
where $r$ is the energy fraction of the electron carried by the 
photon. 

The single spin asymmetry  for semi-inclusive process 
$A^\uparrow + B \rightarrow C+X$ is defined as 
\be
A_N = \frac{d\sigma ^\uparrow \, - \, d\sigma ^\downarrow}
{d\sigma ^\uparrow \, + \, d\sigma ^\downarrow} \label{an}
\ee 
We assume a generalization of CEM expression by taking into account the 
transverse momentum dependence
of the Weizsacker-Williams  (WW) function and gluon distribution function.
The numerator of the SSA can be written as
\bea
\frac{d^{4}\sigma^\uparrow}{dydM^2d^2\bfq_T}-\frac{d^4\sigma^\downarrow}
{dydM^2d^2\bfq_T} &=&
\frac{1}{s}\int [d^2\bfk_{\perp\gamma}d^2\bfk_{\perp g}]
\Delta^{N}f_{g/p^{\uparrow}}(x_{g},\bfk_{\perp g})
f_{\gamma/e}(x_{\gamma},\bfk_{\perp\gamma}) \nonumber\\&&
\times\>\delta^2(\bfk_{\perp\gamma}+\bfk_{\perp g}-\bfq_T)
\hat\sigma_{0}^{\gamma g\rightarrow c\bar{c}}(M^2)
\label{num-ssa}
\eea

where $y$ is the rapidity and $q_T$ in the transverse momentum of the
charmonium; $\Delta^{N}f_{g/p^{\uparrow}}(x_{g},\bfk_{\perp g})$     
is the gluon Sivers function, $ f_{\gamma/e}(x_{\gamma},\bfk_{\perp\gamma})$
is the photon distribution of the electron, given in the WW approximation.  
The denominator would have a similar expression involving the unpolarized   
gluon distribution of the proton; $f_{g/p}(x_g,\bfk_{\perp g})$, for which  
we use a gaussian form of $k_\perp$ distribution 
and a similar gaussian form for the transverse momentum dependence of the WW
function. To extract the asymmetry produced by the Sivers function, we
calculate the weighted asymmetry ~\cite{vogelsang-weight} 
\be
A_N^{\sin({\phi}_{q_T}-\phi_S)} =\frac{\int d\phi_{q_T}
[d\sigma ^\uparrow \, - \, d\sigma ^\downarrow]\sin({\phi}_{q_T}-\phi_S)}
{\int d{\phi}_{q_T}[d{\sigma}^{\uparrow} \, + \, d{\sigma}^{\downarrow}]}
\label{weight-ssa}
\ee

where ${\phi}_{q_T}$ and  $\phi_S$ are the azimuthal angles of
the $J/\psi$ and proton spin respectively.  For the gluon Sivers function we
have used a model in our analysis, which has been  used in the literature   
to calculate SSA in semi-inclusive deep inelastic scattering (SIDIS)
\cite{Anselmino-PRD72} and Drell-Yan (DY) process \cite{Anselmino2009} 
(see \cite{us} for details). The parameters are taken from 
\cite{2011-parmeterization}. Other parameters we use are   

\hspace{3 cm}$\langle{k_{\perp g}^2}\rangle=\langle{k_{\perp \gamma}^2}
\rangle=0.25\>GeV^2$.

%%%%%%%%%%%%%%%%%%%%%%%%%%%%%%%%%%%%%%%%%%%%%%%%%%%%%%%%%%%%%%%%%%%%%%
\section{Numerical results}
%%%%%%%%%%%%%%%%%%%%%%%%%%%%%%%%%%%%%%%%%%%%%%%%%%%%%%%%%%%%%%%%%%%%%%

In this section, we present numerical estimates of the Sivers asymmetry 
in charmonium production in different experiments. 
Model  I refers to the
parametrization in \cite{Anselmino2009} and (a) refers to the
parametrization of the gluon Sivers function in terms of an average of the u
and d quark Sivers function \cite{us}.  
The estimates are obtained using GRV98LO for gluon distribution functions.
We point out that the scale evolution of the TMD's including the Sivers
function has been worked out in \cite{evo1,evo2} and recently it 
has been noted in \cite{evo3} that in SIDIS the evolution indeed affects the   
Sivers asymmetry. However, in the model we consider for charmonium
production, namely the
color evaporation model, the only relevant scale is $M^2$ which is the
invariant mass of the heavy quark pair. This is integrated between a narrow
region, from $4 m_c^2$ to $4 m_D^2$ irrespective of the center-of-mass
energy of the experiment. So the scale evolution of the TMDs is not expected
to affect the asymmetry too much.

In Figs. 1 and 2 we have shown the asymmetry
($A_N^{\sin({\phi}_{q_T}-\phi_S)}$)
as a function of rapidity $y$ and $q_T$ 
respectively for 
COMPASS ($\sqrt s = 17.33$ GeV) and eRHIC ($\sqrt s = 31.6 $ GeV) energies.  
The plots are for model I (a) and (b) for two choices of the gluon Sivers
function. We obtain sizable asymmetry in the kinematical regions of all the
experiments for model I (b). The asymmetry is smaller in model I (a).
For COMPASS as well as for eRHIC the asymmetry increases with
$y$, reaches a maximum and then decreases. This maximum is reached at $y \approx
0.6$ for COMPASS and at $y \approx 1.2$ for eRHIC for model I (b). 
The asymmetry increases with $q_T$ for both models, and for higher values of
$q_T \approx 0.6-0.7$ GeV, it becomes relatively steady.

%%%%%%%%%%%%%%%%%%%%%%%%%%%%%%%%%%%%%%%%%%%%%%%%%%%%%%%%%%%%%%%%%%%%%%%%%%%%%%
\begin{figure}  
% \hspace*{-2cm} 
\includegraphics[width=0.49\linewidth,angle=0]{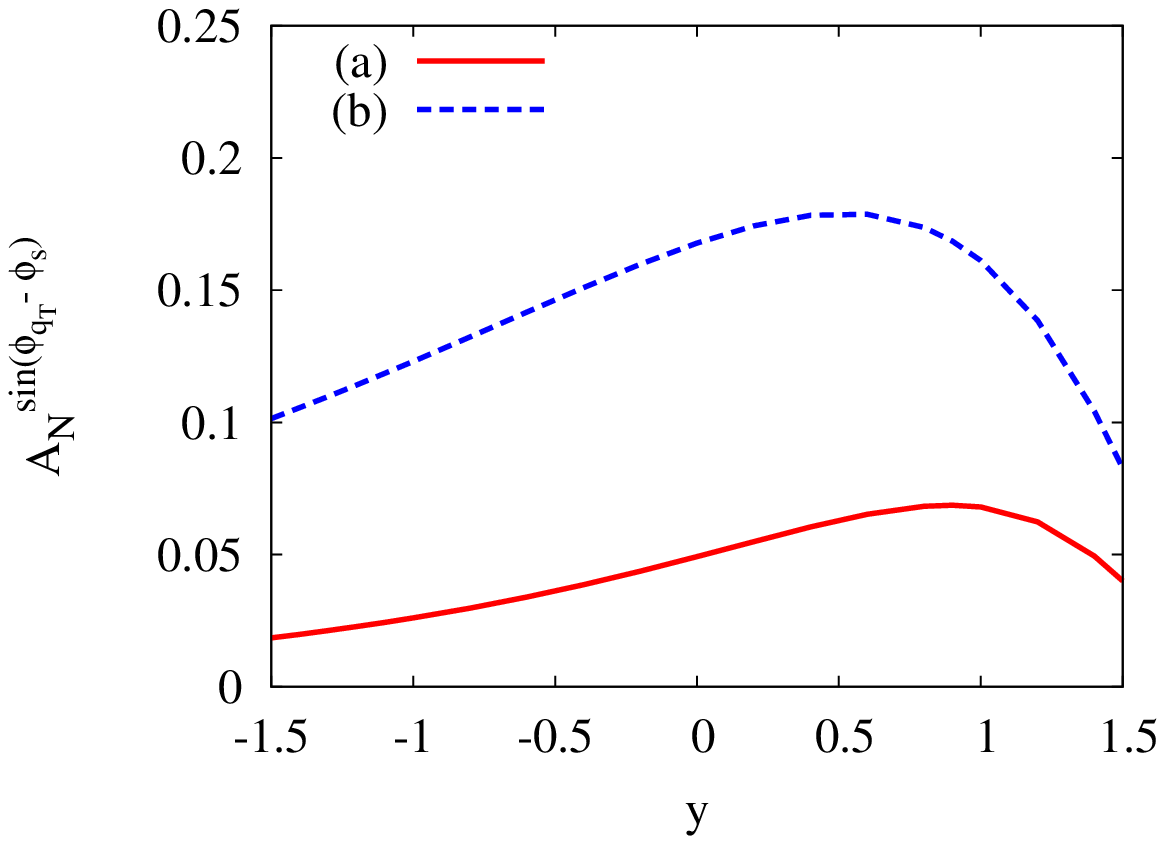}\hspace*{0.2cm}
\includegraphics[width=0.49\linewidth,angle=0]{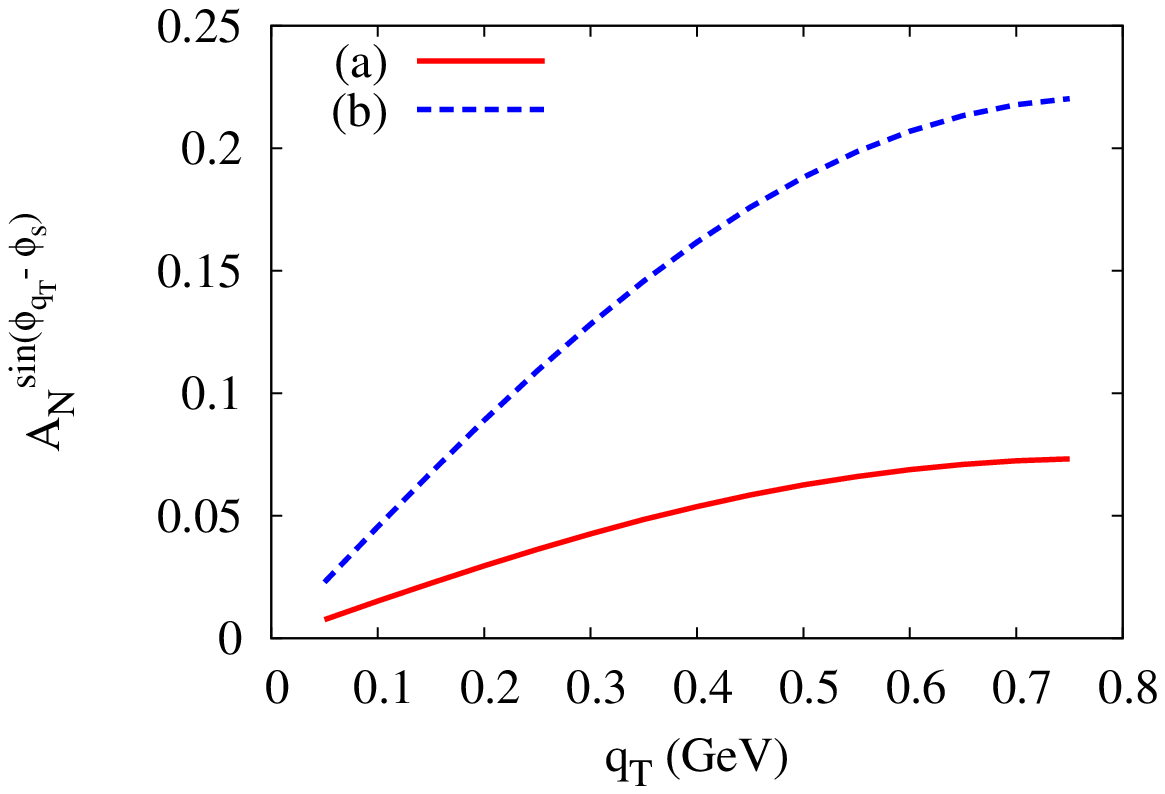}
 \caption{The single spin asymmetry $A_N^{\sin({\phi}_{q_T}-\phi_S)}$   
for the $e + p^\uparrow \rightarrow e+J/\psi+X$ at COMPASS  as a function
of y (left panel) and $q_T$ (right panel). 
The plots are for model I with two parameterizations (a) (solid line)
and (b) 
(dashed line). The integration ranges are $(0 \leq q_T \leq 1)$~GeV and 
$(0 \leq y \leq 1)$. 
The results are given at $\sqrt s = 17.33$~GeV.}
\end{figure}    
%%%%%%%%%%%%%%%%%%%%%%%%%%%%%%%%%%%%%%%%%%%%%%%%%%%%%%%%%%%%%%%%%%%%%%%%%%%%%%

\begin{figure}
% \hspace*{-2cm}
\includegraphics[width=0.49\linewidth,angle=0]{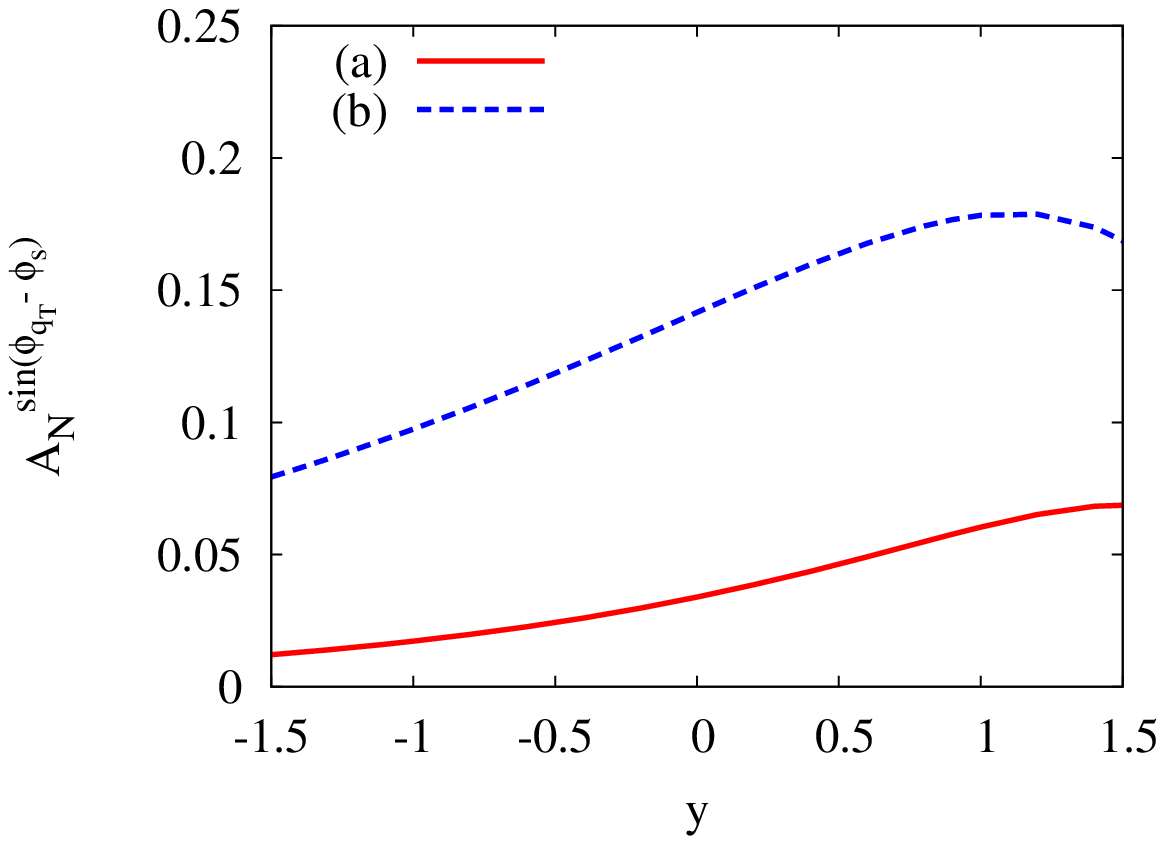}\hspace*{0.2cm}
\includegraphics[width=0.49\linewidth,angle=0]{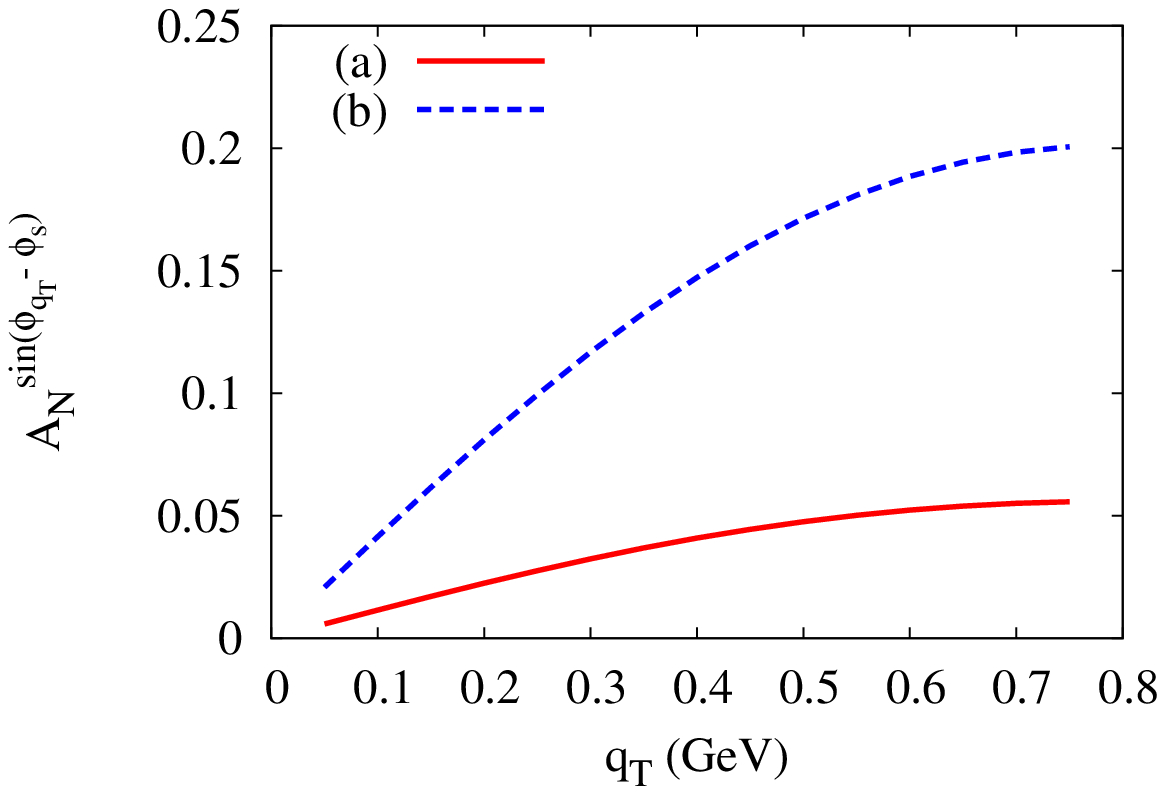}
 \caption{The single spin asymmetry
$A_N^{\sin({\phi}_{q_T}-\phi_S)}$
 for the $e + p^\uparrow \rightarrow e+J/\psi+X$ at eRHIC as a function of y 
(left panel) and $q_T$ (right panel). 
The plots are for model I with two parameterizations (a) (solid line)
and (b)
(dashed line). The integration ranges are $(0 \leq q_T \leq 1)$~GeV and 
$(0 \leq y \leq 1)$. 
 The results are given at $\sqrt s = 31.6$~GeV.}
\end{figure}
%%%%%%%%%%%%%%%%%%%%%%%%%%%%%%%%%%%%%%%%%%%%%%%%%%%%%%%%%%%%%%%%%%%%%%%%%%%%%%%%
\begin{figure}
% \hspace*{-2cm}
\includegraphics[width=0.49\linewidth,angle=0]{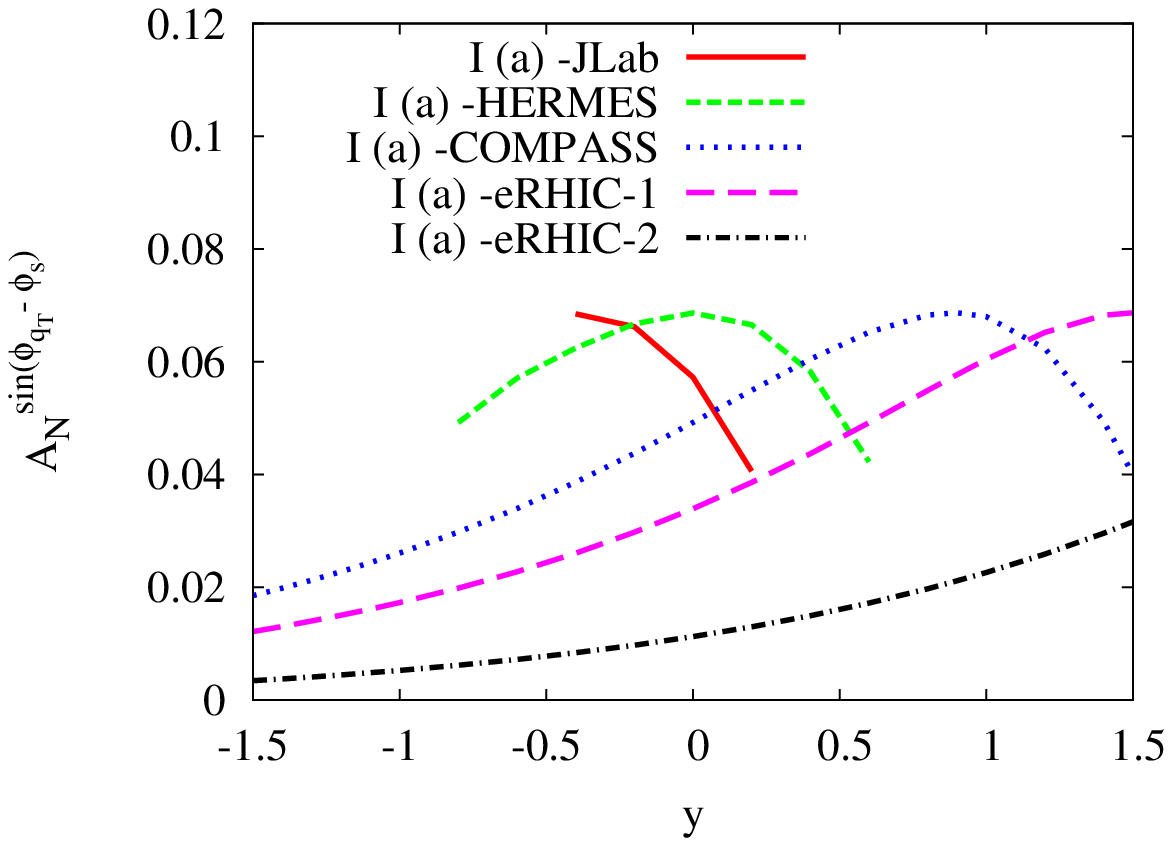}\hspace*{0.2cm}
\includegraphics[width=0.49\linewidth,angle=0]{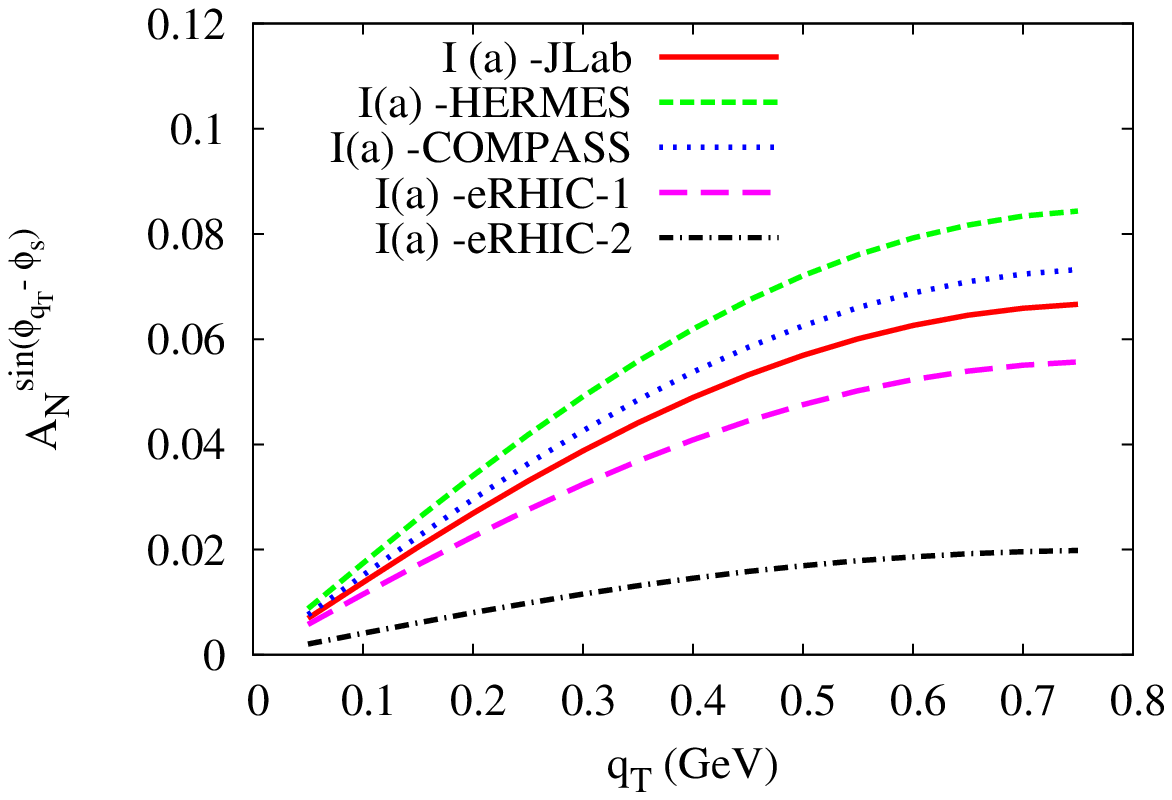}
 \caption{The single spin asymmetry
$A_N^{\sin({\phi}_{q_T}-\phi_S)}$
for the $e + p^\uparrow \rightarrow e+J/\psi+X$ as a function of y (left
panel)  
and $q_T$ (right panel). The plots are for model I with parameterization (a) 
compared for JLab ($\sqrt s = 4.7$~GeV) (solid line), 
HERMES ($\sqrt s = 7.2$~GeV) (dashed line), COMPASS ($\sqrt s =
17.33$~GeV) 
(dotted line), eRHIC-1 ($\sqrt{s}=31.6$~GeV) (long dashed line)
and 
eRHIC-2 ($\sqrt{s}=158.1$~GeV) (dot-dashed black line).}
\end{figure}
%%%%%%%%%%%%%%%%%%%%%%%%%%%%%%%%%%%%%%%%%%%%%%%%%%%%%%%%%%%%%%%%%%%%%%%%%%%%%%%%%

In Fig. 3 we have shown  a comparison of the $y$ and $q_T$ dependence of the
asymmetry at different experiments, namely at JLab, HERMES, COMPASS and
 eRHIC. Different experiments cover 
different kinematical regions, and the plots  clearly show that the 
asymmetry is sizable, and  that  SSA's in charmonium production is a useful
tool to extract information on the gluon Sivers function.

\section*{Acknowledgments}

We thank the organizers of the QCD Evolution Workshop, May 14-17, 
2012, Jefferson Lab, Newport News, Virginia, USA for invitation and support.
This work has been done in collaboration with R. Godbole, A. Misra and 
V. Rawoot.

%\begin{thebibliography}{000} %for 3 digits
%\begin{thebibliography}{00}  %for 2 digits


\begin{thebibliography}{00}    %for 1 digit

\bibitem{tmd-fact1}P.~J.~Mulders and R.~D.~Tangerman,
  %``The Complete tree level result up to order 1/Q for polarized deep
  % inelastic leptop
  Nucl.\ Phys.\ B {\bf 461}, 197 (1996)
  [Erratum-ibid.\ B {\bf 484}, 538 (1997)]
  [hep-ph/9510301];
  %%CITATION = HEP-PH/9510301;%%
D.~Boer and P.~J.~Mulders,
  %``Time reversal odd distribution functions in leptoproduction,''
  Phys.\ Rev.\ D {\bf 57}, 5780 (1998)
  [hep-ph/9711485];
  %%CITATION = HEP-PH/9711485;%%

\bibitem{AdamsBravar1991}   D.~L.~Adams {\it et al.}  [FNAL-E704
Collaboration],
  %``Analyzing power in inclusive pi+ and pi- production at high x(F) with a
  % 200-GeV po
  Phys.\ Lett.\ B {\bf 264}, 462 (1991);
  %%CITATION = PHLTA,B264,462;%%
  A.~Bravar {\it et al.}  [Fermilab E704 Collaboration],
  %``Single spin asymmetries in inclusive charged pion production by
  % transversely polar
  Phys.\ Rev.\ Lett.\  {\bf 77}, 2626 (1996).
  %%CITATION = PRLTA,77,2626;%%


\bibitem{KruegerAllogower1999} K.~Krueger, C.~Allgower, T.~Kasprzyk, 
H.~Spinka, D.~Underwood, A.~Yokosawa, G.~Bunce and H.~Huang {\it et al.},
  %``Large analyzing power in inclusive pi+- production at high x(F) with a
  % 22-GeV/c po
  Phys.\ Lett.\ B {\bf 459}, 412 (1999);
  %%CITATION = PHLTA,B459,412;%%
  C.~E.~Allgower, K.~W.~Krueger, T.~E.~Kasprzyk, H.~M.~Spinka, 
D.~G.~Underwood, A.~Yokosawa, G.~Bunce and H.~Huang {\it et al.},
  %``Measurement of analyzing powers of pi+ and pi- produced on a hydrogen
  % and a carbon
  Phys.\ Rev.\ D {\bf 65}, 092008 (2002).
  %%CITATION = PHRVA,D65,092008;%%
\bibitem{Hermes} A.~Airapetian {\it et al.}  [HERMES Collaboration],
  %``Observation of a single spin azimuthal asymmetry in semiinclusive pion
  % electro pro
  Phys.\ Rev.\ Lett.\  {\bf 84}, 4047 (2000)
  [hep-ex/9910062];
  %%CITATION = HEP-EX/9910062;%%
%   A.~Airapetian {\it et al.}  [HERMES Collaboration],
  %``Single spin azimuthal asymmetries in electroproduction of neutral pions
  % in semiinc
  Phys.\ Rev.\ D {\bf 64}, 097101 (2001)
  [hep-ex/0104005].
  %%CITATION = HEP-EX/0104005;%%

% 4
\bibitem{Compass}   V.~Y.~Alexakhin {\it et al.}  [COMPASS Collaboration],
  %``First measurement of the transverse spin asymmetries of the deuteron in
  % semi-inclu
  Phys.\ Rev.\ Lett.\  {\bf 94}, 202002 (2005)
  [hep-ex/0503002].
  %%CITATION = HEP-EX/0503002;%

\bibitem{alesio-review} 
  U.~D'Alesio and F.~Murgia,
  %``Azimuthal and Single Spin Asymmetries in Hard Scattering Processes,''
  Prog.\ Part.\ Nucl.\ Phys.\  {\bf 61}, 394 (2008)
  [arXiv:0712.4328 [hep-ph]].
  %%CITATION = ARXIV:0712.4328;%%

% 6
\bibitem{Sivers1990} D.~W.~Sivers,
  %``Single Spin Production Asymmetries from the Hard Scattering of
  % Point-Like Constitu
  Phys.\ Rev.\ D {\bf 41}, 83 (1990);
  %%CITATION = PHRVA,D41,83;%%
%  D.~W.~Sivers,
  %``Hard scattering scaling laws for single spin production asymmetries,''
  Phys.\ Rev.\ D {\bf 43}, 261 (1991).
  %%CITATION = PHRVA,D43,261;%%

\bibitem{tmd-fact2} M.~Anselmino, M.~Boglione and F.~Murgia,
  %``Single spin asymmetry for p (polarized) p ---> pi X in perturbative
  % QCD,''
  Phys.\ Lett.\ B {\bf 362}, 164 (1995)
  [hep-ph/9503290];
  %%CITATION = HEP-PH/9503290;%%
  M.~Anselmino and F.~Murgia,   
  %``Single spin asymmetries in transversely polarized proton anti-proton
  % proton inclus
  Phys.\ Lett.\ B {\bf 442}, 470 (1998)
  [hep-ph/9808426];
  %%CITATION = HEP-PH/9808426;%%
  M.~Anselmino, M.~Boglione and F.~Murgia,
  %``Phenomenology of single spin asymmetries in p - polarized p ---> pi
  % X,''
  Phys.\ Rev.\ D {\bf 60}, 054027 (1999)
  [hep-ph/9901442].
  %%CITATION = HEP-PH/9901442;%%

\bibitem{fact} J.~C.~Collins and D.~E.~Soper,
  %``Back-To-Back Jets in QCD,''
  Nucl.\ Phys.\ B {\bf 193}, 381 (1981)
  [Erratum-ibid.\ B {\bf 213}, 545 (1983)]
  [Nucl.\ Phys.\ B {\bf 213}, 545 (1983)];
  %%CITATION = NUPHA,B193,381;%%
X.~-d.~Ji, J.~-p.~Ma and F.~Yuan,
  %``QCD factorization for semi-inclusive deep-inelastic scattering at low
  % transverse m
  Phys.\ Rev.\ D {\bf 71}, 034005 (2005)
  [hep-ph/0404183].
  %%CITATION = HEP-PH/0404183;%%

\bibitem{sing} E.~L.~Berger and D.~L.~Jones,
  %``Inelastic Photoproduction of J/psi and Upsilon by Gluons,''
  Phys.\ Rev.\ D {\bf 23}, 1521 (1981);
  %%CITATION = PHRVA,D23,1521;%% R. Bair, R. R\"uckl, Phys. Lett. 
R.~Baier and R.~Ruckl,
  %``Hadronic Production Of J/Psi And Upsilon: Transverse Momentum
  %Distributions,''
  Phys.\ Lett.\  B {\bf 102}, 364 (1981);
  %%CITATION = PHLTA,B102,364;%%
R.~Baier and R.~Ruckl,
  %``On inelastic leptoproduction of heavy quarkonium states,''
 Nucl.\ Phys.\  B {\bf 201}, 1 (1982).

\bibitem{octet} G. T. Bodwin, E. Braaten, G. P. Lepage, Phys. Rev.
 {\bf D 46}, 1914 (1992). 
% 23
\bibitem{hal} F.~Halzen, Phys.\ Lett.\ B {\bf 69}, 105 (1997);
F.~Halzen and S.~Matsuda,
  %``Hadroproduction of Quark Flavors,''
  Phys.\ Rev.\ D {\bf 17}, 1344 (1978). 
  %%CITATION = PHRVA,D17,1344;%%

% 20  
\bibitem{fri} H. Fritsch, Phys.\ Lett.\ B {\bf 67}, 217 (1977).
% 21
\bibitem{ce2} O.~J.~P.~Eboli, E.~M.~Gregores and F.~Halzen,
  %``Color evaporation description of inelastic photoproduction of $J/\psi$
  % at DESY HER
  Phys.\ Rev.\ D {\bf 67}, 054002 (2003).
  %%CITATION = PHRVA,D67,054002;%%
\bibitem{wwf1}C.~F.~von Weizsacker,
  %``Radiation emitted in collisions of very fast electrons,''
  Z.\ Phys.\  {\bf 88}, 612 (1934).
  %%CITATION = ZEPYA,88,612;%%

% 31
\bibitem{wwf2} E.~J.~Williams,
  %``Nature of the high-energy particles of penetrating radiation and status
  % of ionizat
  Phys.\ Rev.\  {\bf 45}, 729 (1934).
  %%CITATION = PHRVA,45,729;%%


\bibitem{feng} F.~Yuan,
  %``Heavy Quarkonium Production in Single Transverse Polarized High Energy
  % Scattering,
  Phys.\ Rev.\ D {\bf 78}, 014024 (2008)
  [arXiv:0801.4357 [hep-ph]].
  %%CITATION = ARXIV:0801.4357;%%

\bibitem{ww} B.~A.~Kniehl,
  %``Elastic e p scattering and the Weizsacker-Williams approximation,''
  Phys.\ Lett.\ B {\bf 254}, 267 (1991).
  %%CITATION = PHLTA,B254,267

\bibitem{vogelsang-weight} W.~Vogelsang and F.~Yuan,
  %``Single-transverse spin asymmetries: From DIS to hadronic collisions,''
  Phys.\ Rev.\ D {\bf 72}, 054028 (2005)
  [hep-ph/0507266].

\bibitem{Anselmino-PRD72}M.~Anselmino, M.~Boglione, U.~D'Alesio,
 A.~Kotzinian, F.~Murgia and A.~Prokudin,
  %``Extracting the Sivers function from polarized SIDIS data and making
  % predictions,''
  Phys.\ Rev.\ D {\bf 72}, 094007 (2005)
  [Erratum-ibid.\ D {\bf 72}, 099903 (2005)]
  [hep-ph/0507181].

\bibitem{Anselmino2009} M.~Anselmino, M.~Boglione, U.~D'Alesio, S.~Melis, 
F.~Murgia and A.~Prokudin,
  %``Sivers effect in Drell-Yan processes,''
  Phys.\ Rev.\ D {\bf 79}, 054010 (2009)
  arXiv:0901.3078 [hep-ph].

\bibitem{us} R. M. Godbole, A. Misra, A. Mukherjee, V. S. Rawoot, Phys.Rev.
{\bf D 85},
094013 (2012);
arXiv:1201.1066 [hep-ph]
 
 


\bibitem{2011-parmeterization}
M.~Anselmino, M.~Boglione, U.~D'Alesio, S.~Melis, F.~Murgia and
A.~Prokudin,\\
%``Sivers distribution functions and the latest SIDIS data,''
arXiv:1107.4446 [hep-ph].
  %%CITATION = ARXIV:1107.4

\bibitem{evo1} S.~M.~Aybat and T.~C.~Rogers,
  %``TMD Parton Distribution and Fragmentation Functions with QCD
  % Evolution,''
  Phys.\ Rev.\ D {\bf 83}, 114042 (2011);
  arXiv:1101.5057 [hep-ph].
  %%CITATION = ARXIV:1101.5057;%%

% 38
\bibitem{evo2} S.~M.~Aybat, J.~C.~Collins, J.~-W.~Qiu and T.~C.~Rogers,
  %``The QCD Evolution of the Sivers Function,''
  arXiv:1110.6428 [hep-ph].
  %%CITATION = ARXIV:1110.6428;%%
.   
% 39
\bibitem{evo3} S.~M.~Aybat, A.~Prokudin and T.~C.~Rogers,
  %``Calculation of TMD Evolution for Transverse Single Spin Asymmetry
  % Measurements,''
  arXiv:1112.4423 [hep-ph].
  %%CITATION = ARXIV:1112.4423;%%


\end{thebibliography}
\end{document}